# Automatic Recall of Software Lessons Learned for Software Project Managers

Tamer Mohamed Abdellatif[*], Luiz Fernando Capretz, and Danny Ho

*Department of Electrical & Computer Engineering, Western University, 1151 Richmond Street, London, Ontario N6A 5B9, Canada*

**Abstract**

***Context:*** Lessons learned (LL) records constitute the software organization memory of successes and failures. LL are recorded within the organization repository for future reference to optimize planning, gain experience, and elevate market competitiveness. However, manually searching this repository is a daunting task, so it is often disregarded. This can lead to the repetition of previous mistakes or even missing potential opportunities. This, in turn, can negatively affect the organization's profitability and competitiveness.

***Objective:*** We aim to present a novel solution that provides an automatic process to recall relevant LL and to push those LL to project managers. This will dramatically save the time and effort of manually searching the unstructured LL repositories and thus encourage the LL exploitation.

***Method:*** We exploit existing project artifacts to build the LL search queries on-the-fly in order to bypass the tedious manual searching. An empirical case study is conducted to build the automatic LL recall solution and evaluate its effectiveness. The study employs three of the most popular information retrieval models to construct the solution. Furthermore, a real-world dataset of 212 LL records from 30 different software projects is used for validation. Top-k and MAP well-known accuracy metrics are used as well.

***Results:*** Our case study results confirm the effectiveness of the automatic LL recall solution. Also, the results prove the success of using existing project artifacts to dynamically build the search query string. This is supported by a discerning accuracy of about 70% achieved in the case of top-k.

***Conclusion:*** The automatic LL recall solution is valid with high accuracy. It will eliminate the effort needed to manually search the LL repository. Therefore, this will positively encourage project managers to reuse the available LL knowledge – which will avoid old pitfalls and unleash hidden business opportunities.

---

[*] Corresponding author.
E-mail addresses: tmohame7@uwo.ca (T. M. Abdellatif), lcapretz@uwo.ca (L. F. Capretz), danny@nfa-estimation.com (D. Ho)

## 1. Introduction

The dynamic nature of the software industry makes it natural for software practitioners to face a lot of crucial decisions. Those decisions are made during the different stages of the software development lifecycle and on different organizational hierarchical tiers. Decisions vary from developers' implementation decisions, passing through project management, portfolio and release management decisions, and finally reaching high management decisions. The decisions could be either successful or unsuccessful, leading to success stories or failure stories, respectively. Some organizations record those experiences as lessons learned (LL) records and keep them within the organization's repository [1]. Thus, LL repositories can be considered the organization's memory.

The LL could be conceived of as an important part of the organization's memory and accumulative experience and knowledge. LL could be guidelines, handling scenarios or tips related to what went wrong (mistakes) or what went right (opportunities) in certain situations or events. In addition, LL could be a success that the organization wants to repeat, or a failure that the organization wants to avoid in the future. The need to preserve the organization's knowledge, which could be lost due to several reasons, such as expert turnover, calls for the adoption of these LL repositories. The LL concept is evolving, and multiple organizations have their own LL repositories [1].

It is valuable to highlight that LL differ from best practices. In contrast to the best practices that capture only successful scenarios, the LL can capture both success and failure scenarios. Also, best practices are ideas that are recommended on the industrial scope and could be localized to the organization, while LL are organization-oriented and could be globalized to the industrial scope.

LL representation should give information about the problem or opportunity and how to apply the LL recommendations. This information should include the context, the problem/opportunity, and the LL recommendations. The context field describes the situation where the LL is applicable. The problem/opportunity field clarifies the need to apply the LL actions in order to avoid a problem or to leverage an opportunity. Finally, the recommendations field describes the actions that can be followed in order to avoid a problem or to leverage an opportunity. Table 1 shows an LL example in a software company. In this example, the development team should be at the customer premises, so issuing an entry visa for the team can cause a planning issue. For this reason, the LL or the organization recommendation is to plan for this ahead as shown in the LL recommendation section. It is important to highlight that the LL representation can differ from one organization to another. For example, the LL record can be described as a flat text, as in the case of the dataset employed in this paper, without using specific attributes or fields.

Table 1
Lessons Learned Example

| Attribute | Value |
|---|---|
| Context | One of our project constraints is to have the development team onsite (at the customer's site), and our customer is in X country. |
| Problem/Opportunity | Issuing a visitor visa for our team members takes a lot of time especially during high seasons. |
| Recommendations | Try to make your staffing plan updated and covering 1 or 2 months ahead. Try to start the visa issuing process, for any member, 4-5 weeks ahead of the start date of the planned task at the customer's site. Try to seek your customer's support in getting a long-term visa (example: 6 months) with multiple entries. |

*Some sensitive information regarding customer's identity and country was updated or removed due to the non-disclosure agreement*

The problem is that those LL repositories are rarely reviewed or could even be abandoned [2], [3]. This could lead to the repetition of the same mistakes, and could waste the opportunity to benefit from previous success stories. Disregarding LL repositories could be due to the lack of knowledge of relevant LL by project managers (PMs) or due to the need to continuously remind them of the existence of new relevant LL [2]. Although, this can be overcome by manually searching for relevant LL records by PMs, this is effort and time costly, especially when searching in unstructured information. Also, there could be other reasons for disregarding LL repositories, such as time limitation [2]. Dulgerler and Negri [3] underlined that the current traditional LL repositories are overlooked by practitioners. The authors claimed that one of the reasons for this overlooking is attributed to the difficulty of searching this LL repository for relevant LL. This is the same reason for neglecting the LL knowledge from NASA LL system. As reported by Li [4], one of the reasons for overlooking NASA LL system by practitioners and PMs is that it is hard to manually query the system for relevant records [4].

The primary objective of this research is to leverage the benefits that can be gained from the organization's LL knowledge. We believe that this can be achieved by both facilitating the retrieval of relevant information and boosting knowledge about relevant and useful LL. We aim to achieve this by employing information retrieval (IR) techniques to provide adequate LL retrieval classifiers. To the best of our knowledge, we are the first to employ IR techniques in mining the software LL repository; this is consistent with the literature survey conducted by Chen, Thomas and Hassan [5].

Most of the current research work regarding the improvement of LL retrieval, as per our knowledge, relies on techniques such as case-based reasoning. Case-based reasoning depends on describing the LL in a case-based question-answer format which, in turn, requires reformation of the existing LL repository records. This is not the case for our proposed IR-based classifiers, as there is no reformation needed for the LL records.

In addition, we provide a technique to automatically recall relevant LL for PMs. Our proposed solution is automatic as it dynamically facilitates the search of the proposed LL classifiers without the need for manual involvement by PMs. This is achieved by linking different software project management artifacts to yield useful insights. This is achieved by constructing the search queries from the already existing project artifacts, including project management issues and risks, instead of relying on manual search, which has proven to be a waste of time and effort. We have found that project management issue records and risk records can be effectively used as search queries to recall LL records relevant to the project at hand.

We have performed an empirical case study on real industrial projects' datasets in order to evaluate our proposed solution, that is, employing IR models to recall LL that are relevant to these projects. Three of the most popular IR models have been compared in our case study. In addition, we have constructed 88 LL classifiers using a variety of parameter configurations for the IR models considered to investigate the impact of different classifier configurations on the performance results. Those different classifiers have been benchmarked using the same dataset and performance measurements to guarantee a neutralized comparison. We have found that the classifier's performance is very sensitive to the configurations. In terms of LL, the relevance of the retrieved LL records is affected by the different configurations. Also, we have found that some of the employed IR models outperform other models within the context of our dataset and LL retrieval improvement problem.

As described, most of the available research work regarding the improvement of LL awareness relies on case-based reasoning techniques. As mentioned, these techniques require reformatting the existing LL repositories, which is not a practical solution for practitioners, as it necessitates extra effort. In this paper, we propose the employment of IR models in order to improve the awareness of the existing relevant LL records. In our solution, we aim to automatically recall the LL relevant to the in hand project and push them to the PM. In our research, we try to answer three main research questions:

RQ1: Can we automatically, rather than manually, recall and push the relevant LL to PMs using IR-based LL classifiers?

RQ2: Can project artifacts be used to construct on-the-fly queries to recall LL records relevant to the project in hand?

RQ3: Do the configurations of the LL classifiers have an impact on the performance results?

In order to answer these research questions, we have conducted an empirical case study. In this study, as will be described in detail, we constructed multiple LL classifiers based on different techniques using multiple parameter configurations. We aim to get a positive answer to the first research question by achieving convenient performance results from the classifiers considered.

Also, in our case study, we used two types of the available project artifact records, issue records and project risk register records, to dynamically construct the query string on-the-fly and search the constructed classifiers. It is important to clarify that by issue records we mean project management issues, like cost management issues and team management issues, not development issues or bugs.

By using the existing artifacts, we bypassed the need of the users to manually construct the query string, and we provided an automatic search process. The relevant accuracy, or the performance results, of the lists retrieved by the LL classifiers should give us an answer to the second research question.

In addition, the analysis of the results of the different configured classifiers provides an answer to the third research question.

This paper is organized as follows: In Section 2, we provide a background regarding IR techniques and demonstrate the state-of-the-art and the related work of software LL retrieval. We illustrate our proposed solution and the case study methodology in Section 3. Also, in this section, we describe in detail how we validated our proposed solution by conducting an empirical case study, then we demonstrate our results in Section 4. We discuss our findings and link them together to answer our research questions in Section 5. Section 6 defines our research threats to validity. We conclude our research work and propose potential future work in Section 7.

## 2. Background

Detailed in this section is the basic background regarding IR models and software LL recall literature and state-of-the-art.

### 2.1. Information Retrieval Models

IR refers to the process of finding a relevant document or information of interest within a collection of documents or artifacts. In this paper, we use the term information retrieval to refer to text IR in mining software repositories. Usually the information within the searched collection, in the case of IR, is in an unstructured format (i.e., natural language text) [6]. The input to the IR classifier is a query, or question, and the result is a list of the documents relevant to this query [7]. For example, the web search engines, such as Google, can be considered as one of the IR applications where the user provides a query, describing the need or the question, and the search engine tries to answer the user's question by replying with a list of the most relevant web content.

There are multiple IR models that can be used to construct classifiers, and they vary based on their theories, such as simple keyword matching and statistics. There are two main factors which affect the operation and the accuracy of the IR classifier. The first factor is the preprocessing steps, which are employed to process the text inputs before forwarding them to the IR classifier. In our case, the text inputs include both the LL records, which are used to construct the IR classifier, and the issue/risk records, which are used to query the constructed classifier. Different preprocessing steps from the natural language processing literature can be used. Later in Section 3.2.1, we provide some details regarding the preprocessing steps used.

The IR model parameters are the second factor. Each of the IR models or techniques has its own specific parameters which drive the classifier construction and operation. Examples of these parameters can be the *similarity*, the method to calculate the document relevance to the query, and the *term weight*.

In the following subsections, we give an introduction to three of the most popular IR models from the literature which we used in our study. We aimed to consider both algebraic and probabilistic models in our study.

    5

These models are: Vector Space Model (VSM), Latent Semantic Indexing (LSI), and the Latent Dirichlet Allocation (LDA).

### 2.1.1. Vector Space Model

The VSM is an algebraic IR model. VSM relies on representing the documents' corpus in a matrix format of terms versus documents (t x d matrix). In this matrix format, each term in the corpus vocabulary, where the vocabulary contains all the different terms, has a term weight value corresponding to each document in the corpus. The row dimension value of the matrix represents the number of the different terms, where each row represents a term. On the other hand, the column dimension value represents the number of the various documents in the corpus. In each term row, the term has a non-zero weight value if the term exists in the corresponding document, and a zero value otherwise. The term can represent a single word and its weight can be calculated using a term weighting method. In order to decide if two documents, or a document from the corpus and a query, are relevant, the VSM model compares these two documents' columns or vectors from the terms versus the documents' matrix. This comparison is achieved using a configured similarity method which can be, for example, the inner product of the two documents' vectors. To consider that two documents are relevant, they should have one or more common terms. The VSM model returns a proportional continuous similarity value according to the number of common terms between the two compared documents.

The VSM model has two main configurable parameters:

- *Term weight*: the term weight in a document. The basic weight method is the *Boolean* method whose value is '1' if the term appears in the document, and '0' otherwise. Other popular weighting methods are term frequency (*tf*), which is the number of times the term appears in a document, and term frequency-inverse document frequency (*tf-idf*), which is an extended version of the original *tf* with the consideration of the term popularity in corpus documents [6]. For *tf-idf*, the term weight for a certain document is high if it appears with high frequency in this document and, at the same time, the term is rare and has a low frequency within the overall document corpus.
- *Similarity*: the method used to calculate the similarity degree between two document vectors, or, as in our case, between a document and a query. Popular similarity methods include *cosine* distance and *overlap* methods [6].

### 2.1.2. Latent Semantic Indexing

The LSI model is an extension of the VSM model. Unlike VSM, LSI takes the context or topic into consideration instead of only matching the terms which can have different meanings, polysemy, within different topics. For LSI, documents sharing the same topics, even if they do not share the same terms, can be considered similar documents. This is very important in the case of synonymy and polysemy [7], [8]. To achieve this goal, LSI employs a technique called singular value decomposition (SVD). SVD decomposes the term-document matrix (t x d), used by VSM, into three new matrices: the term-topic or T matrix (t x k), the diagonal eigenvalues matrix S (k x k), and the topic-document matrix D (k x d). The k value represents the number of topics, which is a value provided by the model user. The SVD technique works on reducing the rank of both T and D matrices to the provided k value [9]. During this decomposition, the SVD technique works on grouping the co-occurring terms, which appear together, into one topic.

The LSI has three parameters as follows:

- *Term weight*: the same as in the VSM model.
- *Similarity*: the same as in the VSM model.
- *K* or *number of topics*: the number of topics remaining after the SVD reduction.

### *2.1.3. Latent Dirichlet Allocation*

The LDA is a generative probabilistic model [10], [11]. LDA considers the context of terms by eliciting the topics within the documents' corpus. For the LDA model, each document can be composed of one or more topics with a different membership degree for each topic. Also, the topics can be constructed from one or more terms. Each term can belong to one or more topics with a certain membership value [7].

In order to infer which topics exist in which documents, as well as which terms belong to which topics, LDA employs latent variable models, such as Gibbs sampling or Bayesian inference machine learning algorithms [11], in an iterative inference process.

LDA model has several parameters which can be listed as follows [7]:

- $\alpha$: the document-topic smoothing parameter for the probability distribution.
- $\beta$: the term-topic smoothing parameter for the probability distribution.
- *Similarity*: the same as in the VSM model.
- *K* or *number of topics*: the number of topics to be created by the LDA model.
- *Number of iterations*: the number of iterations considered for the inference process.

### *2.2. Related Work*

The literature review conducted by Abdellatif, Capretz, and Ho [12] has shown that much of the available research work that tackles software project artifacts analysis serves the needs of developers, while it merely provides lip-service to managers. This has been the first trigger for our research, which targets managers in order to participate in closing this research gap. Also, the literature review results have shown that only a small portion of the considered studies focused on domains that are related to managers, including incident management and software effort estimation, while the major part of work focused on domains that support developers.

Most of the available LL research focuses on either the LL process or the implementation of a standalone LL repository system [1]. To the best of our knowledge, only a small portion of the available research has tried to facilitate the LL retrieval process.

Harrison [13] has introduced a standalone software LL system. In his implementation, he has tried to improve the usability of the information retrieval by providing different search options. The system provided the ability to search based on domain, keyword, or repository navigation. However, this does not eliminate the need of the users to manually define the search query string. Furthermore, NASA developed an LL repository portal and allowed practitioners to search the system for relevant LL records [14]. Yet, NASA has reported that the system is underutilized and ineffective due to the difficulty of manual searching for relevant records, which obstructed the adoption of the system by NASA practitioners [4]. LessonFlow system [15] is another example of a standalone portal for LL capturing and validation with search capabilities. These solutions are different from our proposed solution as they did not address the need of manual search by users. On the other hand, our solution makes use of the existing project artifacts to search for relevant LL.

Sary and Mackey [16] have introduced RECALL, which is a case-based reasoning (CBR) system. CBR has been employed to improve relevant LL retrieval by users. RECALL work differs from our proposed research in significant ways. First, the employed CBR technique is different from the proposed IR techniques that are presented in this paper. Second, the RECALL system relies on describing the LL in a case-based question-answer format. This format is difficult to follow for the existing organizations' LL repositories.

To the best of our knowledge, the work of Weber et al. [17] is the only available work that does not require users to fully construct the query string. They introduced an LL retrieval tool called "ALDS," and they embedded this tool in a decision-making tool called "HICAP." They provided an implementation for ALDS within the task decomposition of the project planning phase. However, ALDS differs from our proposed solution in multiple ways. First, ALDS employs the CBR technique, while our solution employs the IR technique. IR and CBR are different in some aspects; e.g., in CBR, cases are stored in a "case representation" format, where additional inferred knowledge can be kept to make them better fitting for reasoning and learning in new situations [18]. The second difference is related to LL similarity evaluation; ALDS relies on the indexing of LL in a question-answer format, where users have to go through answering the questions while describing their task condition. In contrast, this limitation is not required for our solution since it relies on automatically querying the LL classifiers or the search engine using data extracted from the project artifacts. The queries, the issue or risk records, are extracted from the existing project artifacts, which are issue/risk register documents. We describe our proposed methodology in more detail in the next section.

It is worth mentioning that IR models have been used to solve several problems in the software engineering domain, such as bug localization [19], [20], [21], concept location [22], and regression tests selection [23], but have not been employed to improve the LL recall as per our knowledge [5]. Thus, to the best of our knowledge, we are the first to employ IR techniques to solve the LL recall issue within the software engineering context [5].

Moreover, the impact of the parameter configurations on the information retrieval classifiers' performance was studied in other software engineering domains, other than LL recall, such as bug localization [19] and equivalent requirements [24]. This was our motive to consider the parameter configurations impact within our study.

## 3. Case Study Methodology

In this case study, our target is to answer our research questions by evaluating the ability of our solution to recall the LL and push them to PMs in an automatic way.

### 3.1. Dataset Collection

One of the most challenging steps for the success of our case study was to collect the dataset needed to evaluate and answer our research questions. Keeping in mind the necessity of confidentiality and competitiveness within the software industry, it was not an easy task to get access to the needed dataset, especially that we targeted real industrial records. After communicating with our industrial network, we successfully received the data needed from an industrial partner, which is a large and reputable multinational software company with a workforce of 800+ employees. Our industrial partner is both ISO 9001 and CMMi Level 3 certified, with more than 20 years in the global IT services domain. The company has global branches all over the world, including North America, Canada, and Arab Gulf countries, such as the Emirates, Saudi Arabia, and Kuwait. The company provides software solutions within seven different industries, including telecommunications, banking, education and government sectors, besides strategic education programs and partnership with multiple Arab Gulf governments, including Dubai and Qatar governments. The data provided is under a non-disclosure agreement. According to this agreement, the dataset records have been made totally anonymous by removing all sensitive data, such as customer names and project names. The collected dataset is two-fold; the first part is the LL repository, while the second part is the project issues/risks register documents.

Table 2
LL Record Sample

| Project ID | Lesson Learned Description |
|---|---|
| Project <id> | There is too much context switching amongst team members.<br>Since this is unavoidable due to attrition, separation, career planning, etc., constant update to organization chart within tools team is needed. Team members are to share domain knowledge, back each other up as part of organization planning. |

The LL repository sample provided contains 212 LL records from 30 different software projects. Each LL record is represented by both the project's identification (ID) number field, identifying the project which has reported the LL, and the description field. The description field contains a description of the LL and its context in a flat text format. Table 2 shows an example of an LL record.

Regarding the project issue/risk records, we have received 55 issue/risk records from five different projects that are different than the 30 projects used for the LL records. Those records acted as the query string for our case study.

Table 3
Issue Records Sample

| Project ID | Issue Description |
|---|---|
| Project <id> | Delay in signing requirement and design documents by client. |
| Project <id> | There is no availability of a technical writer. |
| Project <id> | Additional ramp up effort and constant re-clarification of roles and responsibilities due to context switching. |
| Project <id> | Project contract is not clear and has not been signed yet |

Table 4
Risk Records Sample

| Project ID | Risk Description |
|---|---|
| Project <id> | Source code is at client side with no remote access. If no appropriate backup and labelling are processed, then an issue can happen or code loss can occur. |
| Project <id> | If there is delay in requirement document sign off by customer as planned on <date>, this can lead to delay of schedule and can affect milestone dates and resources travel dates. |
| Project <id> | If roles of different stakeholders are not set clear, then this will impact the scoping and requirements sign off. |
| Project <id> | If there is any issue in issuing an entry Visa for the team leader, then this can lead to delay of schedule and can affect milestone dates and resources travel dates. |

The projects are from different domain verticals, including governmental, educational, and telecommunications projects. Also, the customers of these projects are from different countries. All the dataset records are written in English. Table 3 and Table 4 show samples of both issue and risk records, respectively. In addition, a summary of the collected data is shown in Table 5.

Table 5
Collected Data Summary

| Record Type | Number of Records Collected | Number of Source Projects |
|---|---|---|
| Lesson Learned | 212 | 30 different projects* |
| Issue or risk (query) | 55 | 5 different projects* |

\* The 30 projects from which the LL records were collected are different than the 5 projects from which the queries were collected.

## Gold Set Construction

After receiving this two-part dataset (project artifacts and LL), the next step was to construct our gold set. In order to be able to evaluate any LL classifier, the gold set should contain a mapping set of each query examined and the relevant results expected for this query. In order to construct this map, each of the provided issue/risk records was mapped to the relevant LL records from the LL repository. As this map could be subjective based on the users – practitioners and PMs in our case – of the retrieval model, we adopted a procedure similar to the one recommended by Kitchenham et al. [25] in performing data extraction while conducting a systematic literature review in the case of having a single main researcher. So, the initial mapping was conducted by the main author, the single main researcher in our case who is a subject matter expert (SME). Then, a review meeting was scheduled with another SME from the partner software company. In the review meeting, the company SME reviewed the mapping of the issues/risks to the relevant LL records. In the case of disagreement, the two SMEs held discussions till reaching consensus. After finalizing and agreeing on the whole mapping set, it was baselined. This final mapping set was used for the evaluation and benchmarking of the different LL classifiers within our case study. We summarize the gold set construction process in Fig. 1.  In addition, an example of mapping relevant LL to a query is shown in Table 6.

Table 6
Mapping Relevant Lessons Learned to a Query Example

| Query | Relevant Lessons Learned |
|---|---|
| Additional ramp up effort and constant re-clarification of roles and responsibilities due to context switching. | There is too much context switching amongst team members. Since this is unavoidable due to attrition, separation, career planning, etc., constant update to organization chart within tools team is needed. Team members are to share domain knowledge, back each other up as part of organization planning. |

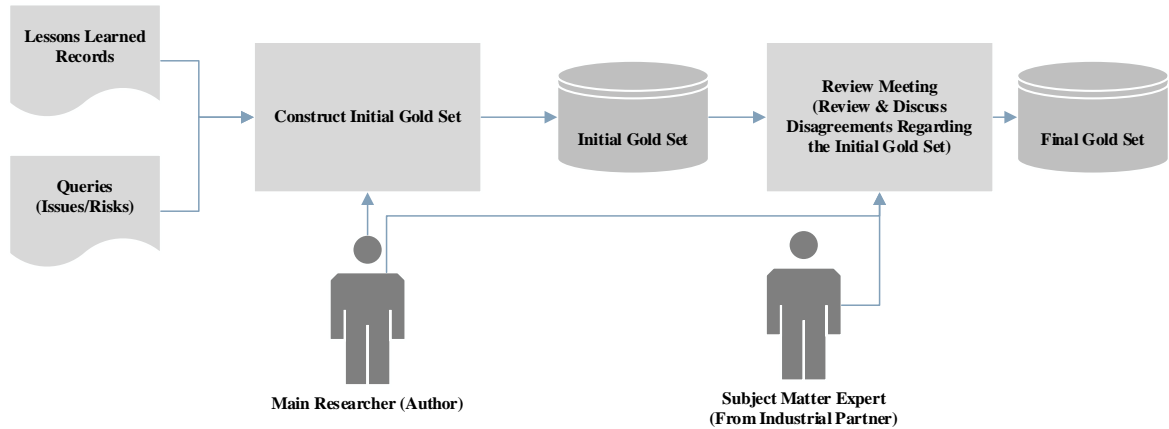

Fig. 1. Gold Set Construction Process

### 3.2. Case Study Design

The following subsections describe our case study design, including the constructed LL classifiers, and the performance metrics employed.

#### 3.2.1. Lessons Learned Classifiers

In our study, we have relied on the IR-based classifiers. We have employed three popular IR models from the literature, namely: LSI, LDA, and VSM. We have had to define three types of configurations: data representation, preprocessing steps and model-based parameter configurations. A summary of the parameter configurations considered is shown in Table 7.

Data Representation Configuration

Both the search query, which is formed of project artifacts (issues/risks), and the LL records have been represented using their full description field values. We have only relied on the description field value as other fields, such as "title," were not defined for the dataset provided. Regarding the queries, we used the text of the issue or risk record as it is, then applied to it the preprocessing steps considered. In addition, for our case study, each considered issue or risk record text is used separately to construct a query that is used to query the LL classifier to retrieve the most relevant LL records to this query.

Table 7
Parameter Configurations

| Parameter | Value |
|---|---|
| *Common parameters* | |
| Preprocessing steps | None |
| | Stemming |
| | Stopping |
| | Stemming and stopping |
| *VSM model parameters* | |
| Term weight | tf-idf |
| | Sublinear tf-idf |
| | Boolean |
| Similarity | Cosine |
| | Overlap |
| *LSI model parameters* | |
| Term weight | tf-idf |
| | Sublinear tf-idf |
| | Boolean |
| Number of topics | 32 |
| | 64 |
| | 128 |
| | 256 |
| Similarity | Cosine |
| *LDA model parameters* | |
| Number of topics | 32 |
| | 64 |
| | 128 |
| | 256 |
| Number of iterations | Until model convergence |
| Similarity | Conditional probability |

*Since LDA smoothing parameters are automatically configured by the tool used, these parameters are not mentioned in this table.*

## Preprocessing Steps Configuration

Since both the documents' corpus, LL repository in our case, and the query comprise unstructured information, they are preprocessed before being forwarded to construct or query the LL classifiers. The preprocessing has a key role in reducing any information noise which can be a source of confusion to the LL classifiers. It is a

   11

common practice from IR research to apply one or more preprocessing steps from the natural language processing (NLP) literature [7]. The preprocessing configuration describes how data (project artifacts and LL records) is preprocessed before being forwarded to the IR algorithm to build the LL classifier. Since the selection of the appropriate preprocessing steps is an open research area [7], we have chosen to employ two of the most common techniques from the literature, namely: *stopping* and *stemming.*

The following is a brief description of the two preprocessing steps which we applied in our study:

- Stopping step: removing the common stop words from the English language, such as "the" and "an". Such words are very common and have high appearance frequency within the document, which can impact the relevance score while not representing a real relevance of the document to the query.
- Stemming step: reducing the words to their morphological roots or stems. For example, "stem" is the root for both "stemming" and "stems". In the experiments, we exploit the Porter's algorithm [26] to perform the stemming step. Since the documents, i.e., LL, issue and risks records, within the dataset considered are not long (see Tables 2, 3 and 4), stemming was employed in order to address the data sparseness.

There is no existence of multi-part identifiers or compound terms in our validation dataset. Thus, there has been no need for other steps such as splitting, which is frequently used in the case of source code preprocessing [7]. Also, other quality check steps, such as abbreviation extension and spelling checks, are performed by the quality assurance team before storing the LL records and project artifacts within the organization's repositories.

In order to apply these two preprocessing steps, we have used the tool provided by Thomas [27]. We have considered the four combinations of applying these two preprocessing steps: not applying any of the two steps (none), applying *stemming* individually, applying *stopping* individually and applying both *stemming* and *stopping.*

### Model-Based Parameter Configuration

For the LSI model, there are three parameters which should be configured: *number of topics*, *term weight,* and *similarity*. Since there is no optimal selection method for the *number of topics*, and since it is still an open research topic, we have considered four values from the literature [19] for the *number of topics*; "32," "64," "128" and "256." Those chosen values should cover the different ranges of the *number of topics* values [28]. Regarding the *term weight*, we have considered three methods from the literature [6], namely: *Boolean*, *tf-idf*, and *sublinear tf-idf* methods. For the *similarity*, the *cosine* similarity method has been employed, as it is the most suitable method from the literature for the LSI model [6], [19]. For the LSI model implementation, we employed the *gensim* open source tool [29].

For the LDA model, we have considered the same *number of topics* values as in LSI. Other parameters, including *sampling iterations number*, *topic-word smoothing*, *document-topic smoothing*, and *similarity*, are automatically optimized by the MALLET tool [30], which we have used for our case study experiments. Also, for our query execution, we used the lucene-lda tool, which is implemented by Thomas [31]. The lucene-lda tool employs the *conditional probability* method for the *similarity*, as it is the most appropriate similarity method for IR applications [19], [32].

Regarding the VSM model, there are two parameters: the *term weight* and the *similarity*. For the *term weight*, we have employed the same methods as in the LSI model. For the *similarity*, we have considered both *cosine* and *overlap* methods from the literature [6]. In the case of VSM, the same *lucene-lda* tool used with LDA is employed to construct and query the LL classifiers.

### Overall Considered Configurations

We apply a *fully factorial design* [19] for our experiment, which means that for our study we consider all the combinations of selected parameter values. So for each parameter, every value considered is examined against all values of all other parameters.

As we have followed a fully factorial experiment design, we have considered all the combinations of the parameter values examined in our case study. This experiment design has yielded **88** LL classifiers; **48** LSI classifiers ((1 project artifacts representation) * (1 LL records representation) * (4 preprocessing combinations) * (4 number of topics values) * (3 term weighting methods) * (1 similarity method)), **16** LDA classifiers ((1 project artifacts representation) * (1 LL records representation) * (4 preprocessing combinations) * (4 number of topics values)), and **24** VSM classifiers ((1 project artifacts representation) * (1 LL records representation) * (4 preprocessing combinations) * (3 term weighting methods) * (2 similarity methods)). We have tested and evaluated all of these classifiers.

### 3.2.2. Evaluation Process

For our evaluation process, we followed the Cranfield evaluation methodology [33]. This methodology is suitable for the empirical evaluation of IR models. For this evaluation method, we relied on the gold set built to evaluate multiple LL classifiers. The evaluation process is conducted based on defined performance metrics. In the next subsection, we introduce the performance metrics selected.

We pursued our evaluation process by applying the data preprocessing steps, following the preprocessing combinations, as described in the "Preprocessing Steps Configuration" section, to the LL repository in order to get different preprocessed versions of the LL repository. After that, we built the LL classifiers based on each of the LL repository versions, and then we repeated this for each of the IR model configuration combinations that we considered in the "Model-Based Parameter Configuration" section (see Fig. 2). It is worth mentioning that only the LL records are used to build the LL classifiers, i.e., the query records (issues and risks) are not used to build the LL classifiers.

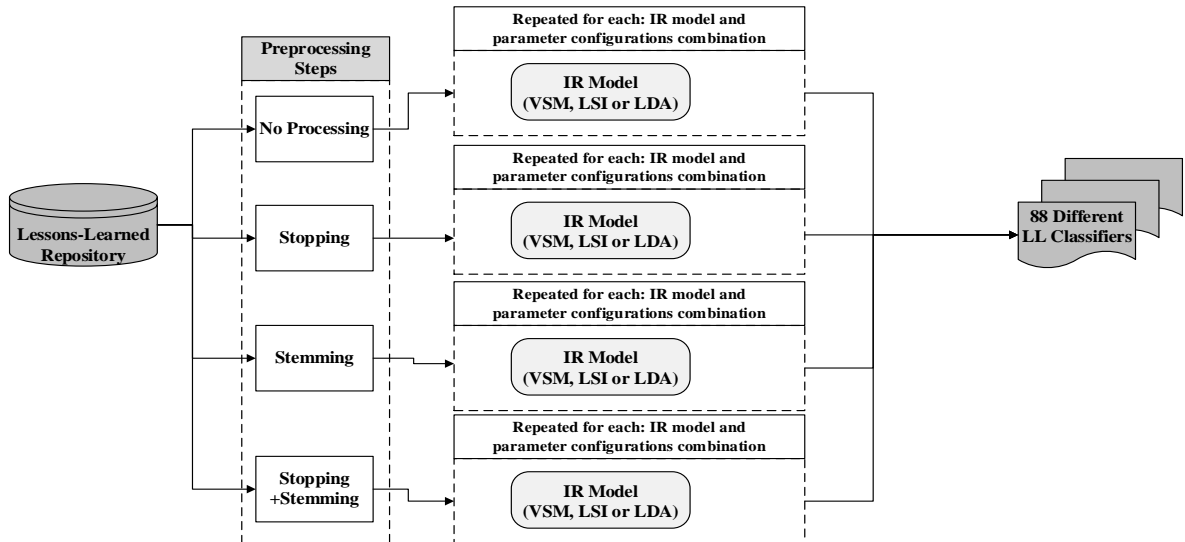

Fig. 2. LL classifiers construction

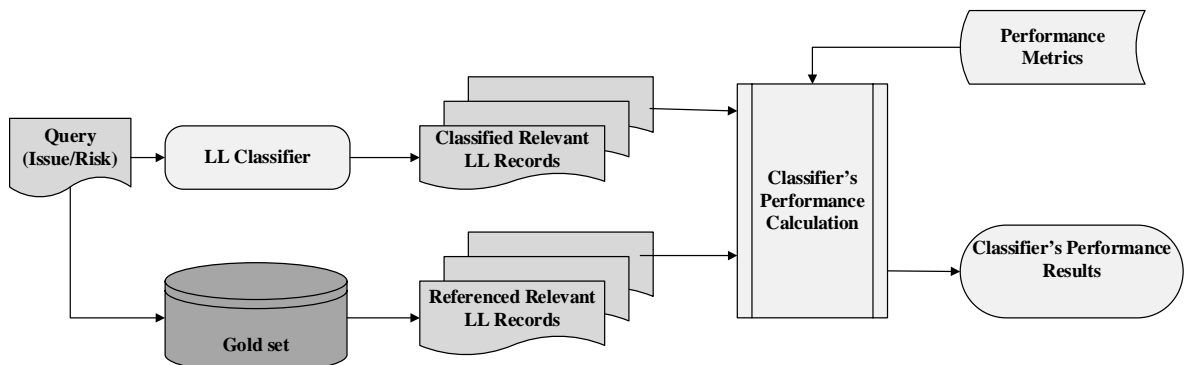

Fig. 3. LL classifier evaluation process and performance results calculation. This process is repeated for each classifier, and is calculated for each of the 55 query results. Then, the average top-20 and MAP are calculated for each classifier

After building the classifiers, we executed each of the queries considered, i.e., issues or risks, using each of these classifiers, and then we recorded the results list. Then, we calculated the performance metrics, as described in the next sections, for each classifier by comparing the results list to the gold set (see Fig. 3).

In addition, we planned to study the impact of the different parameter value configurations, i.e., preprocessing steps and parameter values of models, on the classifiers' performance. In order to conduct this study, we have applied the Tukey's Honestly Significant Difference (HSD) statistical test [34], [35], [36]. The HSD test is a statistical test which has the ability to perform a comparison between different groups in one step. The advantage of the Tukey's HSD test is that it can significantly differentiate between more than two groups based on the statistically significant difference between the groups' mean. The HSD test achieves that by employing the studentized range distribution [34], [35], [36]. For our study, we used the HSD test to statistically compare the impact of the different parameter configurations on the classifier performance. We studied that parameter by parameter.

So, for each parameter (e.g., *term weight*), we compared the different performance results of each parameter value (e.g., *tf-idf* versus *sublinear tf-idf* versus *Boolean*). While studying a certain parameter, the other parameters may vary. The HSD test examines the difference in the mean value between the results of the parameter value pairs. For each of these pairs, if the difference between their mean exceeds the expected standard deviation, then HSD can report these two parameter values as statistically different groups. Therefore, any two parameter values can be either statistically different, i.e., reported as different groups, or not statistically different, i.e., the same group, based on the mean difference. Also, any parameter value can belong to one or more groups.

### 3.2.3. Performance Metrics

To benchmark the performance results for each of the LL classifiers considered, we have employed two of the most popular performance metrics from the literature [6], [7], [19], namely: top-K and MAP (Mean Average Precision).

The top-K accuracy metric calculates the percentage of queries, project issues/risks, in whose top k result records there is at least one LL record relevant to this query, based on the gold set. The top-K value is significant for our case study because it measures the ability of the LL classifier to provide users with at least one relevant result in an advanced position in the results list, which is important to encouraging users to use the new searching tool; this can lead to improvements in the organization's LL recall – our main goal. In our study, we follow the literature by setting the k to 20 in order to measure the accuracy considering the top 20 records from the relevant records retrieved. In the literature [19], [20], the value of 20 was justified as a convenient number of result records through which the user can scroll down before rejecting the search results. Top-K calculations can be formulated as follows [19], [20]:

$$top-K(C_i) = \frac{1}{|Q|} \sum_{j=1}^{|Q|} I(\mathrm{Re}\,levant\,docs(q_j), TopK\,\mathrm{Re}\,cords(C_i, q_j, k)),$$

where $C_i$ is the classifier $i, |Q|$ is the total number of the queries examined, $\mathrm{Re}\,levant\,docs(q_j)$ is a function that returns all of the relevant documents to the *j*th query based on the gold set, $TopK\,\mathrm{Re}\,cords(C_i, q_j, k)$ is a function that returns the top *k* result records from the retrieved list for the $q_j$ by the *i*th classifier $C_i$, and finally $I$ is the intersection function which returns '1' if there is at least one common document between the two document lists returned by the two functions, $\mathrm{Re}\,levant\,docs$ and $TopK\,\mathrm{Re}\,cords$, and returns '0' otherwise.

In our case study, the queries can have more than one relevant LL record, so it is important to measure the ability of the LL classifiers constructed to recall all possible relevant records, as well as evaluate the classifier retrieval precision. In order to fulfill this measurement, we employed the MAP metric, which is one of the most popular and most appropriate metrics, from the literature [6], [33] for this kind of measurement, especially when comparing multiple IR models and with the existence of multiple query sets. The MAP metric can be calculated as the average of the aggregated average precision of each individual query. The MAP equations are formalized by Zhai and Massung [33] as follows:

$$MAP(L) = \frac{1}{m} \sum_{i=1}^{m} avp(L_i),$$

$$avp(L_i) = \frac{1}{|\mathrm{Re}\,l|} \sum_{j=1}^{n} p(j),$$

where $L_i$ is the ranked results list returned by the classifier to answer the $i$th query from the different $m$ queries considered; $avp(L_i)$ is the average precision for the ranked list $L_i$. The $avp$ is calculated for each query, based on the above equation where $p(j)$ is the precision at the ranked record j within the results list $L_i$, $\mathrm{Re}\,l$ is the set of all documents relevant to this query based on the mapping set, the gold set, and $n$ is the count of the records of the results list $L_i$. $p(j)$ is '0' if the jth document is not relevant to the query. Conversely, if the document is relevant to the query, then $p(j)$ will be calculated by dividing the number of relevant documents, identified relevant so far, by the document rank, i.e., $j$ value. For example, if the seventh document within the results list is the fourth relevant document retrieved, then $p(7) = \dfrac{4}{7}$.

## 4. Results

In this section, we describe the results of all the LL classifiers considered. To make it easy to follow, we will discuss the results based on each of the chosen performance metrics, top-K and MAP, separately in the following two subsections. Each subsection starts with the overall discussion of the performance results, then it demonstrates the statistical test results regarding the significant effect of the parameter configurations on the classifier results. Also, we shared the results of all classifiers online as a reference for interested practitioners and researchers [37].

### 4.1.1. Top-K Results

The top-20 performance results regarding the best four classifiers and the worst four classifiers for each of the IR models considered, VSM, LSI, and LDA, are listed in Table 8. When observing the highest performing classifier in each technique, the best top-20 results of 70% were recorded by the VSM and LSI top two classifiers, while the lowest performance was recorded by the LDA top classifier with only 52%. So, the top VSM and LSI models outperformed the top LDA classifier, which is consistent with the literature results for similar problems, such as bug localization [19], [21]. An observation regarding the best two classifiers of VSM and LSI was that both classifiers missed the relevant LL records for almost the same queries (issues/risks) except for only one query. All the missed queries had only three or fewer relevant LL records, which made them hard queries, except for only one missed query which had seven relevant LL records according to the gold set. This indicates that the VSM and LSI best classifiers can be considered good retrieval classifiers for the evaluation dataset in hand.

In addition, the descriptive statistics of the top-20 performance results, in Table 9, demonstrate that the parameter configurations of the LL classifiers had a significant effect on the results. In the case of VSM classifiers, there was a significant difference, about 50% relative improvement (calculated as $\dfrac{70 - 46}{46}$% ), in the performance between the best classifier, 70%, and the worst classifier, 46%, and this could also be observed between the best classifier, 70%, and the median classifier 54%. The same observation was true for the LSI and LDA classifiers, as depicted in Table 9.

In order to statistically study the impact of the configuration values on the performance results, we applied the Tukey's HSD statistical test to the performance results of each of the parameter configuration values. The results of the Tukey's test, regarding the top-20 performance results, illustrated in Table 10 (A), are demonstrated in the following two subsections, in which we use the short term *"performance results"* to refer to the top-20 performance.

Table 8
Lessons Learned Classifiers Performance Results (Best Four and Worst Four Classifiers)

| VSM | | | LSI | | | LDA | | |
|---|---|---|---|---|---|---|---|---|
| Rank | Parameter values | Top-20 | Rank | Parameter values | Top-20 | Rank | Parameter values | Top-20 |
| 1 | Stemming+tf-idf+cosine | 70% | 1 | None+tf-idf+cosine +128 topics | 70% | 1 | Stemming and stopping +32 topic | 52% |
| 2 | Stemming+sublinear tf-idf+cosine | 69% | 2 | None+sublinear tf-idf +cosine+128 topic | 69% | 2 | Stopping+32 topic | 46% |
| 3 | None+sublinear tf-idf+ cosine | 61% | 3 | Stemming+sublinear tf-idf+ cosine+256 topic | 69% | 3 | Stemming and stopping +64 topic | 46% |
| 4 | Stemming and stopping+ sublinear tf-idf+cosine | 61% | 4 | None+tf-idf+cosine+ 256 topic | 69% | 4 | None+32 topic | 41% |
| 21 | Stemming+tf-idf+overlap | 52% | 45 | Stemming+boolean+ cosine+64 topic | 50% | 13 | Stemming and stopping +128 topic | 26% |
| 22 | Stemming+boolean+ overlap | 50% | 46 | None+boolean+cosine +128 topic | 48% | 14 | Stemming+256 topic | 22% |
| 23 | None+boolean+cosine | 46% | 47 | None+boolean+cosine +64 topic | 44% | 15 | None+256 topic | 19% |
| 24 | None+boolean+overlap | 46% | 48 | None+boolean+cosine +32 topic | 43% | 16 | Stemming and stopping +256 topic | 19% |
| Rank | Parameter values | MAP | Rank | Parameter values | MAP | Rank | Parameter values | MAP |
| 1 | Stemming and stopping+ sublinear tf-idf+cosine | 0.189 | 1 | Stemming and stopping+ sublinear tf-idf+ cosine+128 topic | 0.198 | 1 | Stemming+32 topic | 0.096 |
| 2 | Stemming and stopping+ tf-idf+cosine | 0.188 | 2 | Stemming and stopping+ tf-idf+cosine+128 topic | 0.198 | 2 | Stemming and stopping+32 topic | 0.089 |
| 3 | Stemming+tf-idf+cosine | 0.156 | 3 | Stopping+tf-idf+ cosine+64 topic | 0.194 | 3 | None+32 topic | 0.082 |
| 4 | Stemming+sublinear tf-idf+cosine | 0.153 | 4 | Stopping+sublinear tf-idf+cosine+64 topic | 0.194 | 4 | Stopping+32 topic | 0.075 |
| 21 | None+tf-idf+overlap | 0.099 | 45 | None+boolean+cosine+ 128 topic | 0.107 | 13 | Stemming and stopping+128 topic | 0.040 |
| 22 | None+sublinear tf-idf+ overlap | 0.095 | 46 | None+boolean+cosine+ 64 topic | 0.096 | 14 | Stopping+64 topic | 0.036 |
| 23 | None+boolean+cosine | 0.082 | 47 | Stemming+boolean+ cosine+32 topic | 0.086 | 15 | None+256 topic | 0.031 |
| 24 | None+boolean+overlap | 0.081 | 48 | None+boolean+cosine+ 32 topic | 0.085 | 16 | Stemming and stopping+256 topic | 0.030 |

Table 9
Information Retrieval Classifiers Performance Results Descriptive Statistics

| | VSM | | | LSI | | | LDA | |
|---|---|---|---|---|---|---|---|---|
| | Top-20 (%) | MAP | | Top-20 (%) | MAP | | Top-20 (%) | MAP |
| Minimum | 46 | 0.081 | Minimum | 43 | 0.085 | Minimum | 19 | 0.030 |
| 1st Quartile | 52 | 0.111 | 1st Quartile | 55 | 0.132 | 1st Quartile | 26 | 0.043 |
| Mean | 56 | 0.126 | Mean | 59 | 0.153 | Mean | 33 | 0.058 |
| Median | 54 | 0.122 | Median | 59 | 0.163 | Median | 35 | 0.057 |
| 3rd Quartile | 58 | 0.142 | 3rd Quartile | 65 | 0.172 | 3rd Quartile | 41 | 0.065 |
| Maximum | 70 | 0.189 | Maximum | 70 | 0.198 | Maximum | 52 | 0.096 |

*Top-20 results are rounded to two digits, while MAP results are rounded to three decimal points*

In addition to studying the parameter configuration impact, we extended the statistical test by employing the Wilcoxon statistical test to compare the top performer classifier using the preprocessing methods, i.e., stemming and stopping, to the top performer classifier without using any of the preprocessing methods. This was conducted for each of the models considered. The results are discussed later in this section.

<u>Lessons Learned Classifier Parameters</u>

For the VSM classifiers, the HSD test results showed a significant difference in the performance results when using the *cosine* similarity method versus the results of using the *overlap* method. This means that the similarity method employed had an impact on the performance results for the dataset considered in this case study. The *cosine* similarity method showed the best performance results and came in the top group. On the other hand, the *overlap* method results came in the bottom group.

Regarding the VSM classifiers, the test results showed that there was no statistically significant difference when changing the *term weight* parameter value between *tf-idf*, *sublinear tf-idf*, and *Boolean* weighting methods.

For the LSI classifiers, the statistical test showed that the *term weight* parameter had a statistically significant impact on the performance results. Both the *tf-idf* and *sublinear tf-idf* weighting methods came in the top group and had the highest top-20 performance results, while the *Boolean* weighting method came in the bottom group with the lowest performance results.

An overall observation, regarding the *term weight* parameter, was that the *tf-idf* weighting method always showed the highest performance results for both the VSM and LSI models, followed by the *sublinear tf-idf* method, although there was no statistical significance for VSM as described, which is consistent with the results from other IR application studies [19].

The HSD test revealed that the *number of topic*s parameter had a statistically significant impact on the classifiers' performance results. This means that the performance results differed when the classifiers were configured with different topic numbers. This applied for both the LSI and LDA classifiers. However, for LSI, the largest numbers of topics, "128" and "256," came in the top group. This indicates that the more topics used, the better the performance results. On the other hand, for the LDA classifiers, the situation was different, where the smallest numbers of topics, "32" and "64," came in the top groups.

Table 10
Tukey's HSD Statistical Test Results

| VSM | | | LSI | | | LDA | | |
|---|---|---|---|---|---|---|---|---|
| Group | Mean (%) | Preprocessing steps | Group | Mean (%) | Preprocessing steps | Group | Mean (%) | Preprocessing steps |
| A | 59 | Stemming and stopping | A | 60 | Stopping | A | 36 | Stemming and stopping |
| A | 58 | Stemming | A | 60 | Stemming and stopping | A | 34 | Stopping |
| A | 53 | None | A | 60 | Stemming | A | 32 | None |
| A | 53 | Stopping | A | 58 | None | A | 32 | Stemming |
| Group | Mean (%) | Similarity | Group | Mean (%) | Number of topics | Group | Mean (%) | Number of topics |
| A | 58 | Cosine | A | 63 | 128 | A | 45 | 32 |
| B | 53 | Overlap | A | 61 | 256 | AB | 37 | 64 |
| | | | AB | 60 | 64 | B | 28 | 128 |
| | | | B | 54 | 32 | B | 24 | 256 |
| Group | Mean (%) | Term weight | Group | Mean (%) | Term weight | | | |
| A | 58 | tf-idf | A | 63 | tf-idf | | | |
| A | 57 | Sublinear tf-idf | A | 63 | Sublinear tf-idf | | | |
| A | 52 | Boolean | B | 53 | Boolean | | | |

(A) TOP-20 RESULTS

| VSM | | | LSI | | | LDA | | |
|---|---|---|---|---|---|---|---|---|
| Group | Mean | Preprocessing steps | Group | Mean | Preprocessing steps | Group | Mean | Preprocessing steps |
| A | 0.159 | Stemming and stopping | A | 0.170 | Stemming and stopping | A | 0.067 | Stemming |
| B | 0.126 | Stemming | A | 0.164 | Stopping | A | 0.056 | Stemming and stopping |
| B | 0.117 | Stopping | AB | 0.147 | Stemming | A | 0.055 | None |
| B | 0.102 | None | B | 0.132 | None | A | 0.053 | Stopping |
| Group | Mean | Similarity | Group | Mean | Number of topics | Group | Mean | Number of topics |
| A | 0.135 | Cosine | A | 0.161 | 128 | A | 0.085 | 32 |
| A | 0.117 | Overlap | A | 0.161 | 64 | B | 0.053 | 64 |
| | | | A | 0.158 | 256 | B | 0.051 | 128 |
| | | | A | 0.133 | 32 | B | 0.042 | 256 |
| Group | Mean | Term weight | Group | Mean | Term weight | | | |
| A | 0.136 | tf-idf | A | 0.167 | Sublinear tf-idf | | | |
| A | 0.134 | Sublinear tf-idf | A | 0.166 | tf-idf | | | |
| A | 0.109 | Boolean | B | 0.127 | Boolean | | | |

(B) MAP RESULTS

Preprocessing Steps

Table 10 (A) illustrates the HSD test results of applying the four preprocessing combinations on the classifiers' top-20 performance, where there was no statistically significant difference in the results when applying any of the preprocessing steps. This was the case for all the IR models considered, VSM, LSI, and LDA, within the context of the dataset in hand. However, applying both *stemming* and *stopping* together, in the case of VSM and LDA, and applying only *stopping*, in the case of LSI, gave the highest top-20 performance.

Top Performer Classifiers

As clarified, the Wilcoxon test was employed to compare the top performer classifier using the preprocessing methods to the top performer using no preprocessing. This was conducted in three pairs: (1) the top performer using the stemming method versus the top performer using none of the preprocessing steps, (2) the top performer using the stopping method versus the top performer using none of the preprocessing steps, (3) the top performer using both stemming and stopping methods versus the top performer using none of the preprocessing steps. This was conducted for the three models considered and the results are shown in Table 11. The results showed significant difference, i.e., p-value < 0.05, only in two cases for the VSM model where the stemming or stopping methods were employed, while no significant difference was recorded in the other cases including the LSI and LDA classifiers.

Table 11
Top Performer Classifiers Statistical Test Results (Top-20)

| VSM | | LSI | | LDA | |
|---|---|---|---|---|---|
| Top Performer Classifier | p-value | Top Performer Classifier | p-value | Top Performer Classifier | p-value |
| Stemming vs None | 0.025 | Stemming vs None | 0.655 | Stemming vs None | 1 |
| Stopping vs None | 0.046 | Stopping vs None | 0.317 | Stopping vs None | 0.366 |
| Stemming + stopping vs None | 1 | Stemming + stopping vs None | 0.705 | Stemming + stopping vs None | 0.083 |

*4.1.2. MAP Results*

Table 8 lists the MAP performance results regarding the best four classifiers and the worst four classifiers for each of the IR models considered. After analyzing the MAP results, we concluded that some of the insights from the top-20 results still applied. When looking at the top performing classifiers in each model, the highest MAP result of 0.198 was recorded by the top classifier in LSI, followed by 0.189 in VSM, which is similar to the top-20 results. These MAP performance results are satisfactory when compared to other studies from the literature [21], [38]. Also, as in the top-20 results, the LDA top classifier achieved the lowest performance of 0.096, compared to the top performing classifiers in VSM and LSI. In addition, the worst results for both the VSM and LSI classifiers, 0.081 and 0.085, respectively, were slightly different from the LDA top classifier result of 0.096. So, again, our MAP results came aligned with both the top-20 results, from our case study, and the literature results, which provides evidence of the superiority of both VSM and LSI classifiers results over LDA classifiers in different empirical studies [21].

Similar to the top-20 results, the descriptive analysis of MAP performance results, presented in Table 9, indicates that the classifier configuration had a remarkable impact on the performance. This could be inferred from the difference between the VSM best classifier performance of 0.189 and the VSM worst classifier performance of 0.081, which represented more than 100% relative improvement. Also, there was a high difference between the median VSM classifier, 0.122, and the minimum VSM classifier. The same insight applied for both the LSI and LDA results.

In the following subsections, we demonstrate the HSD statistical test results, listed in Table 10 (B), regarding the significant effect of the LL classifiers' configuration on the MAP performance results. Also, the Wilcoxon statistical test results are demonstrated. We refer to the MAP performance results as "*performance results*" in the following two subsections.

## Lessons Learned Classifier Parameters

In the case of the VSM classifiers, the Tukey's test results demonstrated that there was no significant impact of the classifier parameter values on the performance results. This means that, within both the context of our case study dataset and the conducted experiments, neither the *similarity* parameter nor the *term weight* parameter affected the performance of the VSM classifiers.

This is not exactly the same for the LSI classifiers, where the statistical test results revealed the significant impact of the *term weight* parameter on the classifier performance results. The *sublinear tf-idf* term weighting method recorded the highest mean performance value, 0.167, and shared the top group with the *tf-idf* method, while the *Boolean* method came in the bottom group. On the other hand, the statistical test of the *number of topics* parameter demonstrated no significant difference in the performance results.

For the LDA classifiers, a significant difference in the *number of topics* parameter results was reported by the statistical test. The top group comprised the performance results of the "32" topic classifiers, while the bottom group involved the performance results corresponding to "64," "128" and "256" topic configuration values.

## Preprocessing Steps

Regarding the *preprocessing steps* parameter's impact on the performance results, the HSD test showed the significant impact of this parameter on both the VSM and LSI classifiers. Applying both the *stemming* and *stopping* steps together showed the highest mean value for the MAP performance and came in the top groups for both the VSM and LSI classifiers. On the other hand, applying none of the preprocessing steps showed the lowest mean value and came in the bottom groups for both the VSM and LSI classifiers.

In the case of the VSM classifiers, the statistical test classified applying both the *stemming* and *stopping* together in the top group, while the application of other preprocessing steps, including *stemming* alone, *stopping* alone, and using none of the preprocessing steps, came in the bottom group.

For the LSI classifiers, both preprocessing steps configurations of applying the *stemming* and *stopping* steps together, and just the *stopping* step were ranked in the top groups. The *stemming* step was ranked in the middle, and not applying any step came in the bottom group.

Regarding the LDA classifiers, the statistical test inferred no significant impact for the preprocessing steps on the classifiers' performance results.

## Top Performer Classifiers

As in the case of top-20, we compared the top performer classifier using the preprocessing steps to the top performer classifier using none of the preprocessing methods. The Wilcoxon statistical test results are shown in Table 12. The results showed significant difference in the case of VSM while using both stemming and stopping together, and also while using only stopping. Also, there was a significant difference in the case of LSI while using the stopping method. The results showed no significant difference for the rest of the cases.

In the following section, we elaborate on our results analysis and provide our overall findings and observations. Also, we link these findings to our original research questions.

Table 12
Top Performer Classifiers Statistical Test Results (MAP)

| VSM | | LSI | | LDA | |
|---|---|---|---|---|---|
| Top Performer Classifier | p-value | Top Performer Classifier | p-value | Top Performer Classifier | p-value |
| Stemming vs None | 0.469 | Stemming vs None | 0.158 | Stemming vs None | 0.303 |
| Stopping vs None | 0.025 | Stopping vs None | 0.008 | Stopping vs None | 0.889 |
| Stemming + stopping vs None | 0.034 | Stemming + stopping vs None | 0.145 | Stemming + stopping vs None | 0.196 |

## 5. Discussion

In this section, we provide an overall discussion and demonstrate our overall findings from the results of our case study. Also, we provide additional insights by conducting an extension study of the performance of Document to Vector (doc2vec) state-of-the-art model.

Regarding our research questions, the conclusions are based on the analysis of the performance results of the 88 different LL classifiers considered in our case study. We sum up our conclusions as follows:

- In order to have clearer interpretation of the achieved performance results, 70% for top-20 and 0.198 for MAP, an additional investigation of both the precision and recall measurements was done for the top performing models. For the top performer model based on top-20, the precision@10 was 9% and recall@10 was 23%. For the top performer model based on MAP, the precision@10 was 11% and recall@10 was 28%. For the LL retrieval context, users or PMs are more interested in getting relevant results, i.e., recall, which can support their decision making rather than the precision of the items within the retrieved list. Given that fact, the average recall@10 of 28%, even though not a high number relatively, can be considered an adequate result to convince PMs to adopt an LL retrieval model which can provide them with an average of 28% of relevant historical experiences, i.e., LL, within the first 10 retrieved items, which can save them the cost of repeating past mistakes or missing opportunities in the case of overlooking the LL repository. This confirms the effectiveness of employing IR techniques in order to automatically push the relevant LL information to the PMs within organizations. With this convenient level of performance, practitioners can be encouraged to rely on the LL IR-based classifiers to automatically search, within the existing organization's LL repositories, for relevant solutions regarding their in hand issues/risks; this answers our first research question *RQ1*.
- Relying on the available artifacts, such as project management issue and risk registers, that are associated with software development and project management processes, to replace the manual querying of the organization's repositories can be significant. This is a positive answer to our second research question *RQ2*, which is supported by our case study results. Since there is no manual querying needed, the practitioners can explore the organization's repositories without worrying about the burden of manually searching the unstructured data, which can be time and effort consuming. This search process can even be automated.
- Regarding the hypothesis of the impact of the classifier configuration on performance, this was generally found to be significant. The same IR technique showed different performance results considering different configurations, and this provided an answer to our third research question *RQ3*.
- For our study, VSM and LSI IR techniques achieved the best top-20 and MAP performance, followed by LDA.
- Our statistical test of the impact of applying different preprocessing steps showed no significant difference for the top-20 performance results. This can be attributed to our dataset and models. However, since the statistical tests of the impact of applying different preprocessing steps showed significance in our MAP results, for VSM and LSI, and in other cases from the literature, such as bug localization [19], we advise considering those different preprocessing steps in future studies.

An overall observation is that the worst VSM and LSI classifiers' performance results, 46% and 43%, respectively, for the top-20, are slightly lower than the best LDA classifier's performance of 52%. Also, the worst LDA classifier performance, 19%, is significantly far from the worst classifiers in the case of VSM and LSI of 46% and 43%, respectively. The same insight can be inferred from the MAP performance results. This can be considered an indication that the LDA technique is not suitable for the LL recall problem. This indication can be useful for practitioners and researchers who plan to work on similar problems in the future. Also, we advise the consideration of employing the *tf-idf* or *sublinear tf-idf* weighting method together with the *cosine* similarity method, as this combination showed the best classifiers' top-20 and MAP results for both the VSM and LSI techniques.

Since our results indicated that the configurations and the selected IR techniques do matter, we recommend considering different configurations and IR techniques, and to be careful when deciding on the LL classifier to be applied to the problem and dataset in hand.

### 5.1. Document to Vector (Doc2vec) Extension Study

Document to vector (doc2vec), as named in the *gensim* implementation [29], is a new state-of-the-art algorithm, which was initially introduced by Le and Mikolov in 2014[39]. doc2vec overcomes part of the weaknesses of bag-of-words (BOW), the most common text vector representation that is used by some of topic model algorithms such as VSM. doc2vec has two main advantages over BOW, which are both consideration of the words order and the words semantics. doc2vec can operate in one of two model versions, namely the Distributed Memory Model of Paragraph Vectors (PV-DM) and Distributed Bag of Words version of Paragraph Vector (PV-DBOW). The PV-DM model relies on concatenating the paragraph (i.e., document) vector representation and the vector representation of the words to predict the following word within the same context. On the other hand, PV-DBOW relies only on the paragraph vector to predict the next word. doc2vec employs the gradient descent and backpropagation (same algorithms used for the neural network training) in both paragraph and words vectors training [39].

Since doc2vec has shown promising results for similar problems from the software engineering literature [39], extending our primary experiments with the doc2vec experiments within the LL retrieval context could provide useful insights and valuable information to the software engineering research community. For this extension, both DM and DBOW were considered. In addition, we tried to cover different ranges of window sizes, i.e., the distance considered between the current word and predicted word, of 2, 5, 10, 20, 30 and 40. Also, the experiments were repeated for each of the considered preprocessing steps combination from Section 3.2.

The DBOW results are very low for both top-20 and MAP with the highest achieved results of 13% top-20, in the case of no preprocessing or applying the stemming method, and 0.046 MAP, in the case of no preprocessing. This could be attributed to the short size of the dataset documents. On the other hand, the DM results are better, but not satisfactory, with a maximum achieved performance of 52% top-20, in the cases of applying the stemming preprocessing step with 30 window size and applying the stopping step with 5 window size, and 0.1 MAP in the case of applying the stemming preprocessing step with 10 window size. Since doc2vec show more satisfying results, from the literature, than the results from this extension study, a profounder study of the doc2vec performance in the LL retrieval context could be considered for future work. Also, we shared the results of all classifiers online as a reference for interested practitioners and researchers [37].

## 6. Threats to Validity

In this section, we discuss two validity threats regarding the gold set, as well as the dataset representation or context.

*Gold set validity*. In our study, we have relied in our classifier validation on the collection constructed of the queries-relevant LL records mapping. As this mapping collection can be subjective and may cause a threat to the validity of our case study and conclusions, we have taken two mitigation steps. First, as a trial to eliminate any bias, we involved two practitioners in the discussion and construction of this mapping collection. Second, after reaching a consensus from the two practitioners regarding this mapping collection, the collection was baselined. So, even if the collection has any flaw, such as positive or negative falses, the baseline guarantees that the same collection is used to evaluate all the classifiers considered using all the three IR techniques. So, the classifiers were evaluated under the same comparison factors and within the same context.

*Dataset representation*. Although in our empirical study we were keen to consider a significant dataset, including both significant LL records and query records, the dataset considered does not represent the whole of the LL records in the world or even the organization. In addition, we were limited to the dataset provided by our industrial partner, which was out of our control because of data confidentiality restrictions. As this is a common challenge in the context of empirical studies seeking real industrial data, we did our best to come up with solid conclusions by including LL and queries from a variety of projects, domains, and regions. Due to this limitation in the dataset representation, the results and conclusions are not necessarily valid for other contexts. Although our experiment cannot be reproducible, since we cannot share our dataset due to the non-disclosure agreement limitation, we provided the details of our case study design to encourage researchers and practitioners to proceed with similar methodologies and case studies regarding their different datasets.

## 7. Conclusion

Improving the awareness of software organizations' LL records can reform decision making and project management processes. Providing an automatic process to support PMs in obtaining relevant LL records can improve the PMs' awareness of the organization historical experiences. This is crucial to leverage any potential opportunities or to mitigate any previous mistakes. In our paper, we proposed a new automatic LL recall solution. In our solution, we employed the IR techniques for the first time within the software LL retrieval context.

In order to evaluate the effectiveness of our proposed solution, we sought for answers to our research questions by conducting an empirical case study on a real dataset of industrial software projects. In our case study, we considered three state-of-the-art IR techniques, VSM, LSI and LDA, as well as the existing project artifacts, including the project issue and risk records. In addition, we have statistically tested and studied, using the Tukey's statistical test, the impact of considering different LL classifier parameter configurations on the classifiers' performance results. The impact of applying different preprocessing steps on the data records before constructing the LL classifiers was studied as well.

The case study results confirmed the effectiveness of the proposed solution and its ability to provide PMs with relevant LL in an automatic way and, thus, to eliminate the burden of time and effort required to manually get LL. The summary of our main findings is as follows:

- The best top-20 performance result of 70% and MAP performance result of 0.198 were recorded for VSM and LSI classifiers, while LDA classifiers came next.
- Regarding the top-20, the best performance was recorded in the case of the VSM classifier configured using *tf-idf* for the *term weight*, *cosine* for the *similarity*, and *stemming* for the *preprocessing steps* of the LL and the queries. For the best LSI classifier, the configuration was the same for both *term weight* and *similarity* parameters, there was no *preprocessing steps* for the data records, and the *number of topics* was set to "128."
- Regarding the MAP performance results, the best classifiers for both VSM and LSI were achieved using *sublinear tf-idf* for the *term weight*, *cosine* for the *similarity*, *stemming and stopping* for the *preprocessing* *steps*, as well as setting the *number of topics* to "128" for the LSI classifier.

- The statistical analysis of the different classifier configurations indicated the high impact of the configurations on the classifiers' performance. This was elicited from the significant difference between the performance of the best configured classifiers and the worst classifiers. As an example, for the VSM classifiers, the relative improvement between the best and worst classifiers was about 50% for the top-20 and more than 100% for MAP.

As this is the first empirical study to consider applying IR techniques to tackle the automatic retrieval of software LL records, our results represent a value added to the state-of–the-art knowledge, and they can guide interested researchers, practitioners and organizations through the context of LL automatic pushing to PMs.

Since our work is the first, to the best of our knowledge, to apply IR techniques within the context of software LL retrieval, there are many promising opportunities to extend our research. This can be achieved by considering other state-of-the-art IR ranking functions and models, such as Pivoted Length Normalization VSM [40], BM25F [41], [42], and BM25+ [43]. Furthermore, other weighting and preprocessing techniques from the software literature can be employed, such as assigning different weights for different Part of Speech (POS) tags as in [24], and employing the lemmatization technique to perform the data preprocessing. We can also analyze the natural language patterns within the LL and project artifacts, and examine if the patterns can be used to improve the matching performance. Moreover, optimizing the selection of the appropriate IR model configurations, based on the dataset and problem at hand, can be examined. Arguably, this is an open research topic, especially regarding the optimization of the LDA model configurations [44].

In addition, the impact of combining multiple LL classifiers on the performance can be investigated. Different combination techniques from the literature, such as Borda Count [45] and classifier scores addition [46], can also be evaluated.

Finally, we can consider a utility study of the system usage to evaluate the adoption of practitioners for our LL recall solution.

## Acknowledgments

The authors would like to thank their industrial partner for providing the dataset, which is necessary for the evaluation process of this work. Also, they thank the practitioner who helped in constructing, reviewing, and baselining the gold set. This research has been partially funded by the Natural Sciences and Engineering Research Council (NSERC) of Canada through grant 1033906 within the Discovery Program. Also, the first author was awarded an Ontario Graduate Scholarship (OGS). The authors are solely responsible for the results, opinions, and methods presented in the paper, thus they represent neither NSERC nor OGS.